\documentstyle[12pt]{article}

\newcommand{\PSbox}[3]{\mbox{\rule{0in}{#3}\includegraphics{#1}\hspace{#2}}}
\newcommand{\beq}{\begin{eqnarray}}
\newcommand{\eeq}{\end{eqnarray}}
\def\beq{\begin{equation} }
\def\eeq{\end{equation} }
\def\beqa{\begin{eqnarray} }
\def\eeqa{\end{eqnarray} }

\def\non{\nonumber }
\def\ov{\over }

\def\href#1#2{#2}

\begin{document}
\begin{titlepage}
\begin{center}
\vspace*{-1cm} \today \hfill UCB-PTH-99/04\\
hep-th/9902067 \hfill LBNL-42813\\
{}~{} \hfill CERN-TH/99-18\\

\vskip 1cm

{\Large{\bf
Supergravity Models for $3+1$ Dimensional QCD }}

\vskip 1.0cm

{\large Csaba Cs\'aki$^{a,b,}$\footnote{Research fellow, Miller Institute
for
Basic
Research in Science.},\hskip .3 cm Jorge Russo$^{c}$,\hskip .3 cm
Konstadinos Sfetsos$^d$\hskip .3 cm and John
Terning$^{a,b}$}

\vskip 1.0cm

{\it ${}^a$ Department of Physics

University of California, Berkeley, CA 94720, USA

\medskip

${}^b$ Theoretical Physics Group

Ernest Orlando Lawrence Berkeley National Laboratory

University of California, Berkeley, CA 94720, USA
\medskip

${}^c$ Departamento de F\' \i sica, Universidad de Buenos Aires,

Ciudad Universitaria, Pabell\' on I, 1428 Buenos Aires.
\medskip

${}^d$ Theory Division, CERN

CH-1211 Geneva 23, Switzerland}

\vfil

{\tt ccsaki@lbl.gov
,  russo@df.uba.ar
\\ sfetsos@mail.cern.ch
, terning@alvin.lbl.gov 
}

\begin{abstract}

The most general black M5-brane solution of eleven-dimensional
supergravity (with a flat ${\bf R^4}$ spacetime in the brane and a regular
horizon) is characterized by charge, mass and two
angular momenta.
We use this metric to construct general dual
models of large-$N$ QCD (at strong coupling)
that depend on two free parameters.
The mass spectrum of scalar particles
is determined analytically (in the WKB approximation) and numerically
in the whole two-dimensional parameter space.
We compare the mass spectrum with analogous
results from lattice calculations, and
find that the supergravity predictions are close to the
lattice
results everywhere on the two dimensional parameter space except along a
special line. We also examine the mass spectrum of the supergravity
Kaluza--Klein (KK)
modes and
find that the KK modes along the compact D-brane coordinate decouple from
the spectrum for large angular momenta. There are however KK modes charged
under a
$U(1)\times U(1)$ global symmetry which do not decouple anywhere on the
parameter space.
General formulas for the string tension and action are also given.

\phantom{\cite{WittenAdsThermal,GO98,COOT98,jevic,russo,haoz,CORT98,minahan}
}

\end{abstract}
\end{center}
\end{titlepage}

\newpage

\section{Introduction}
\setcounter{equation}{0}
\setcounter{figure}{0}
\setcounter{table}{0}

The conjectured dualities between gauge and
string theories \cite{mal} have been recently exploited
in \cite{WittenAdsThermal}-\cite{minahan} to
construct and investigate models of pure QCD in 3+1 dimensions,
whose main component is the black M5-brane solution of eleven-dimensional
supergravity, which near the branes corresponds to an anti-de Sitter (AdS)
space.
The no-hair theorem implies that the most general model
of this kind that can be constructed (i.e. based on a regular geometry with
M5-brane charge) is obtained from a rotating black M5-brane parameterized by
its
charge, mass and two angular momenta.
The scope of this paper is to calculate the mass spectrum of scalar modes of
this general model in the supergravity approximation,
and study its behavior in the parameter space.
The parameter space is four dimensional,
but the mass parameter can be set to 1 by a choice of mass units;
the charge is related to the 't Hooft coupling $\lambda=g^2 N$
(where $g$ is the Yang--Mills coupling  and $N$ is the number of branes).
It is assumed that $\lambda$ is very large so that the radius of curvature
is much larger than the string scale; this is necessary for
supergravity to be a good approximation to string theory (M theory).
In this regime glueball masses are independent of $\lambda $,
so what remains is a two-dimensional
space spanned by the angular momentum parameters.
When one of the angular momenta vanishes, the model reduces to the
one angular momentum model
examined in Refs. \cite{russo,CORT98}. In our investigation we will use
both
analytic methods (within the WKB approximation, as developed in Refs.
\cite{minahan,RS}) as well as numerical ones based on Ref. \cite{COOT98}.

The static M5-brane has an $SO(5)$ symmetry associated with the internal
$S^4$.
Turning on the angular momentum parameters, this symmetry group
breaks down to the Cartan
subgroup as
$SO(5)\to SO(2)\times SO(2) \sim U(1)\times U(1)$.
The spectrum of the supergravity  field fluctuations can be organized
in
representations
of $SO(5)$ or $SO(2)\times SO(2)$.
The proposal of Refs. \cite{WittenAdsThermal}-\cite{COOT98}
is to identify the $SO(5)$-singlet modes propagating on the Minkowski
boundary of the spacetime  with large-$N$ QCD glueballs.  The dilaton modes
correspond to
$J^{PC}=0^{++}$  glueballs ($J$, $P$, and $C$ being the spin, parity and
charge
conjugation quantum numbers).
In non-supersymmetric, pure $SU(N)$ Yang--Mills theory, there is no
counterpart of the $SO(5)$ global symmetry, so one would expect
that at weak Yang--Mills coupling those Kaluza--Klein (KK)
particles which
transform non-trivially under this group are very massive
and decouple.
This problem was studied for ${\rm QCD}_3$ in \cite{ORT}
where it was shown that the first correction (beyond the $\lambda=\infty$
limit)
to the masses of these states does not lead to their decoupling
in the case of vanishing angular momenta.
A general study for ${\rm QCD}_3$ supergravity models
with three angular momenta was recently given in \cite{RS}.
In this paper, using both analytic (within the WKB approximation)
and numerical methods, we calculate the spectrum of glueballs and of
KK states.
We find that the KK modes on $S^4$ do not decouple in the large $\lambda$
regime
in any region of the two dimensional parameter space (within the
supergravity
approximation). In contrast, the KK modes on the circle
associated with the compact Euclidean time (on the M5-brane worldvolume)
decouple in the limit of large angular
momentum.

Some interesting effects concerning thermodynamical aspects of rotating
D-branes have been recently pointed out in Refs. \cite{gubser}-\cite{caisoh}.
Here we will be considering the slightly different construction of Refs.
\cite{WittenAdsThermal,russo} for
zero-temperature QCD, where the Euclidean time
parameterizes an internal circle, and the Minkowski time is one of
the brane volume coordinates.

\section{The Supergravity Model}
\setcounter{figure}{0}
\setcounter{table}{0}
\setcounter{equation}{0}

\subsection{The metric}

The maximal number of angular momentum parameters for
the rotating M5-brane (dictated
by the rank of the $SO(5)$ isometry group of
rotations of the static M5-brane) is equal to two.
This metric was constructed in \cite{cvet}, though
the expression given there contains a few
minor mistakes which we correct below.
The metric of the rotating M5-brane is given by\footnote{
This differs from Eq.~(12) of \cite{cvet} in the expression for
$\Delta $ (called $f_D^{-1}$ there),
the power of $r$ in the components
$g_{t \varphi_1}, g_{t \varphi_2} $, $g_{\varphi_1 \varphi_2}$,
and a factor $\sin^2\psi $ in $g_{t \varphi_1}$.}
\begin{eqnarray}
ds_{11}^2=&&\hspace*{-0.5cm}
f^{-\frac{1}{3}}(-hdt^2+dx_1^2+\ldots +dx_5^2)+
\frac{f^{\frac{2}{3}}}{\tilde h}dr^2 +f^{\frac{2}{3}}r^2 \big[
(1+\frac{l_1^2 \cos^2 \theta}{r^2}+ \nonumber \\
&&\hspace*{-0.5cm} \frac{l_2^2 \sin^2 \theta \sin^2 \psi}{r^2}
)d\theta^2 +(1+\frac{l_2^2\cos^2\psi}{r^2}) \cos^2\theta d\psi^2 -
2\frac{l_2^2}{r^2}\cos\theta\sin\theta\cos\psi\sin\psi d\theta d\psi
\nonumber \\ && \hspace*{-0.5cm}
-4m\frac{\cosh \delta}{r^5 \Delta f} dt
(l_1\sin^2\theta d\varphi_1+l_2\cos^2\theta \sin^2\psi
d\varphi_2)+\frac{4ml_1l_2\cos^2 \theta\sin^2\theta\sin^2\psi}{r^5\Delta
f}
d\varphi_1d\varphi_2\nonumber \\ && \hspace*{-0.5cm}
+\sin^2 \theta (1+\frac{l_1^2}{r^2}+
\frac{2ml_1^2\sin^2\theta}{r^5\Delta f})d\varphi_1^2+
\cos^2\theta\sin^2\psi (1+\frac{l_2^2}{r^2}+\frac{2ml_2^2\cos^2
\theta\sin^2\psi}{r^5\Delta f} )d\varphi_2^2 \Big], \nonumber \\
\label{uuu}
\end{eqnarray}
where
\begin{equation}
\Delta = 1+\frac{l_1^2}{r^2}\cos^2\theta +\frac{l_2^2}{r^2}(\sin^2\theta
+\cos^2\theta\cos^2\psi )+\frac{l_1^2l_2^2}{r^4}\cos^2\theta\cos^2\psi\ ,
\end{equation}
\begin{equation}
f=1+\frac{2m\sinh^2 \alpha}{\Delta r^3}\ ,
\label{zfff}
\end{equation}
\begin{equation}
h=1-\frac{2m}{\Delta r^3}\ ,
\end{equation}
\begin{equation}
\tilde
h=\frac{1+\frac{l_1^2}{r^2}+\frac{l_2^2}{r^2}+\frac{l_1^2l_2^2}{r^4}-
\frac{2m}{r^3}}{\Delta}\ .
\end{equation}
The horizon is located at $r=r_H$, where $r_H$ is the largest real root of
\beq
\big( r^2 +l_1^2\big)\big( r^2 +l_2^2\big) -2m r =0\ .
\eeq
One can obtain the following
formulas for the ADM mass, entropy and angular momentum:

\beqa
M_{\rm ADM}&=&{V_5V(\Omega_4)\ov 4\pi G_N}\ 2m (1+{3\ov 4}\sinh^2\alpha )\
,\
\ \ \ \ V(\Omega_4)={8\pi^2\over 3}\ ,
\label{madm}\\
S &=& {V_5V(\Omega_4)\ov 4 G_N}\ 2m\ r_H \cosh\alpha \ ,
\label{entro}\\
J_{1,2} &=&{V_5 V(\Omega_4)\over 4\pi G_N}\ m\ l_{1,2}\cosh\alpha \ ,
\label{jjgg}
\eeqa
\beq
G_N={\kappa_{11}^2\ov 8\pi}=2^4\pi^7l_{\rm P}^9\ ,
\label{ggn}
\eeq
where $G_N$ is Newton's constant in 11 dimensions, and $l_{\rm P}$ is
the 11 dimensional Planck length.
The parameter $\alpha$ is related to the (magnetic) charge $N$
and $m$ by
\beq
\sinh^2\alpha = {1\over 2}
\left(\sqrt{(\pi N l_{\rm P}^3/m)^2+1} - 1\right)\ .
\label{sinhalpha}
\eeq
The Hawking temperature and angular velocities are given by
\beq
T_H={3r_H^4+(l_1^2+l_2^2) r_H^2- l_1^2l_2^2 \over
8\pi m r_H^2 \cosh\alpha }\ ,\ \ \ \ \
\Omega_{1,2}={l_{1,2} \over \cosh\alpha\ (r_H^2+l_{1,2}^2 ) }\ .
\label{ppo}
\eeq
These quantities satisfy the first law of black hole thermodynamics:
\beq
dM_{\rm ADM}= T_H dS+ \Omega_1 dJ_1 +\Omega_2 dJ_2 \ .
\label{fll}
\eeq

Let us now go to Euclidean space $\tau= -it$, $l_{1,2}\to il_{1,2}$, and
take the field theory limit as in \cite{mal,russo}:
\begin{equation}
r=U^2l_{\rm P}^3,\ \ \ \ 2m=U_0^6l_{\rm P}^9, \ \ \ \
l_{1,2}=a_{1,2}^2l_{\rm P}^3 \ ,
\end{equation}
so that
$ 2m\sinh ^2 \alpha \to \pi Nl_{\rm P}^3 $.
We obtain the metric
\begin{eqnarray}
ds_{11}^2& = &\frac{\Delta^{\frac{1}{3}}U^2l_{\rm P}^2}{(\pi
N)^{\frac{1}{3}}}
\Big[ (1-\frac{U_0^6}{U^6\Delta})d\tau^2+dx_1^2+\ldots dx_5^2 \Big]+
l_{\rm P}^2\frac{\Delta^{\frac{1}{3}}(\pi N)^{\frac{2}{3}}4dU^2}{U^2 [
(1-\frac{a_1^4}{U^4})(1-\frac{a_2^4}{U^4})-\frac{U_0^6}{U^6}]}+
\nonumber \\ && \frac{l_{\rm P}^2 (\pi
N)^{\frac{2}{3}}}{\Delta^{\frac{2}{3}}}
\Big[ \Delta_1 d\theta^2 +\Delta_2 \cos^2\theta d\psi^2
+2\frac{a_2^4}{U^4}
\cos\theta\sin\theta\cos\psi\sin\psi d\theta d\psi \nonumber \\
&& -\frac{2U_0^3}{U^4(\pi N)^{\frac{1}{2}}} (a_1^2\sin^2\theta d\tau
d\varphi_1+a_2^2\cos^2\theta \sin^2\psi d\tau d\varphi_2)+\sin^2
\theta (1-\frac{a_1^4}{U^4})d\varphi_1^2+\nonumber \\ &&
\cos^2\theta\sin^2\psi (1-\frac{a_2^4}{U^4})d\varphi_2^2 \Big],
\end{eqnarray}
where
\begin{equation}
\Delta_1=
1-\frac{a_1^4\cos^2\theta}{U^4}-\frac{a_2^4\sin^2\theta\sin^2\psi}{U^4}\ ,
\end{equation}
\begin{equation}
\Delta_2=1-\frac{a_2^4\cos^2\psi}{U^4}\ ,
\end{equation}
\begin{equation}
\Delta=1-\frac{a_1^4\cos^2\theta}{U^4}-\frac{a_2^4(\sin^2\theta+
\cos^2\theta\cos^2\psi )}{U^4}+\frac{a_1^4a_2^4\cos^2\theta \cos^2
\psi}{U^8}\ .
\end{equation}
Note that the component $g_ {\varphi_1 \varphi_2}$ vanishes
in the field theory limit, and so do the last terms
in $g_{\varphi_1\varphi_1}$ and
$g_{\varphi_2\varphi_2}$.

The coordinate $\tau $ describes a circle of radius $R_0$, where
$R_0$ is related to the Hawking temperature $T_H$ by
$R_0=(2\pi T_H)^{-1}$, with
\begin{equation}
T_H= {3u_0\over 2\pi A}\ ,\ \ \ \
\label{teeem}
\end{equation}
\begin{equation}
A=\frac{u_H^4u_0^4}{u_H^8-\frac{1}{3}(a_1^4+a_2^4)u_H^4
-\frac{1}{3}a_1^4a_2^4}= {3 u_0^4 u_H^2\ov
(u_H^2-u_{IH}^2)(u_H^2-u_1^2)(u_H^2-u_2^2)}\ ,
\label{aaa1}
\end{equation}
where we have introduced the coordinate $u$ by $U=2(\pi N)^{1/2} u$, and
rescaled $a_{1,2}\to $ \break $2(\pi N)^{1/2}a_{1,2} $.
The constants $u_H^2,u_{IH}^2,u_1^2,u_2^2$
represent the four different solutions for $u^2$ of the equation
\begin{equation}
(u^4-a_1^4)(u^4- a_2^4)-u_0^6u^2=0 ~.
\label{jjj}
\end{equation}
There are
two positive ($u_H^2$, $u_{IH}^2$, $u_H^2>u_{IH}^2$),
and two negative (or complex) solutions ($u_1^2, u_2^2$), with $u_H^2$ and
$u^2_{IH}$
representing the outer and inner horizons respectively.
When $a_1=a_2=a$, the equation simplifies to
\begin{equation}
u^4-a^4=\pm u_0^3u ~,
\label{jjz}
\end{equation}
where the signs $\pm $ corresponds to the inner and outer horizons.
{}From Eq.~(\ref{jjz}) one sees that
when $a\gg u_0$ the two positive solutions get closer
to each other, thus the inner horizon approaches the outer
horizon.

The gauge coupling $g_4^2$ in the $3+1$ dimensional Yang--Mills theory is
given by the ratio between the periods
of the eleven-dimensional coordinates $x_5$
and $\tau $, i.e.
\begin{equation}
\tau=R_0\theta_2, \ \ \ x_5={g_4^2\over 2\pi }R_0\theta_1=
\frac{\lambda}{N} R_0 \theta_1\ ,\ \ \ \
\ \theta_{1,2}=\theta_{1,2}+2\pi\ ,
\end{equation}
where $\lambda \equiv {g_4^2N\over 2\pi}$ is the 't Hooft coupling.
Dimensional reduction in $\theta_1$ gives the type IIA metric
representing the field theory limit
of the rotating D4-brane metric with two angular momentum parameters:
\begin{eqnarray}
ds_{\rm IIA}^2 &=&\frac{2\pi \lambda A}{3u_0}u\Delta^{\frac{1}{2}}
\big[ 4u^2(-dx_0^2+dx_1^2+dx_2^2+dx_3^2)+\frac{4A^2}{9u_0^2}
u^2(1-\frac{u_0^6}{u^6\Delta})d\theta_2^2+\nonumber \\
&& + \frac{4du^2}{u^2((1-\frac{a_1^4}{u^4})(1-
\frac{a_2^4}{u^4})-\frac{u_0^6}{u^6})}+
\frac{\Delta_1}{\Delta}d\theta^2+\frac{\Delta_2}{\Delta}\cos^2\theta
d\psi^2
\nonumber \\ &&
+2\frac{a_2^4}{u^4\Delta }
\cos\theta\sin\theta\cos\psi\sin\psi d\theta d\psi
+ \sin^2\theta \frac{(1-\frac{a_1^4}{u^4})}{\Delta}d\varphi_1^2+
\cos^2\theta\sin^2\psi \frac{(1-\frac{a_2^4}{u^4})}{\Delta}d\varphi_2^2
\nonumber \\ &&
-\frac{4Au_0^2}{3u^4\Delta}(a_1^2\sin^2\theta d\theta_2d\varphi_1+
a_2^2\cos^2\theta \sin^2\psi d\theta_2 d\varphi_2)\Big]\ ,
\label{aaaa}
\end{eqnarray}
where the dilaton field is given by
\begin{equation}
e^{2\Phi}=\frac{8\pi A^3\lambda^3u^3\Delta^{\frac{1}{2}}}{27u_0^3N^2}\ .
\end{equation}
In these coordinates, the metric is independent of $N$, and the
string coupling is of order $1/N$, as expected.
The 't Hooft coupling $\lambda $ appears as an overall factor
of the metric.
For $u_0\neq 0$, curvature invariants have a
finite value at the horizon, and they are suppressed by
inverse powers of
$\lambda $.

The metric (\ref{aaaa}) has a $U(1)^3$ isometry associated with
translations in $\theta_2,\varphi_1, \varphi_2 $.
This should appear as a global symmetry in the corresponding
dual Yang--Mills theory. Since the pure $SU(N)$ QCD has no such
symmetries, one may expect that
states which have charges with respect to $U(1)^3$ have a large mass
compared to the glueball masses. In Section 4 we
calculate the different mass spectra and investigate this
possibility.

\subsection{String tension and action}

The string tension is given by $1/2\pi $ times
the coefficient of $\sum dx_i^2 $, evaluated at the
horizon,
at the angles where it takes its minimum value
\cite{WittenAdsThermal,russo}.
This follows by minimizing the Nambu--Goto action of the string
configuration. The absolute minimum occurs at $\theta=\psi=0$ or $\pi$.
We obtain
\beq
\sigma ={4\ov 3}\lambda A u_0^2\ .
\label{sstt}
\eeq
String excitations should have masses of order
$\sigma^{1/2}$.
The spin $\leq 2$ glueballs that remain in the supergravity approximation
--~whose masses are determined from the Laplace equation~--
have masses which are independent of $\lambda $.

In the field theory limit, the free energy $F$ ($= {\rm Action}\times T_H$)
takes the simple form
\beq\
F=E-T_HS- \Omega_1 J_1-\Omega_2 J_2=-{V_5  \over 3 \pi^3}\ N^3\ u_0^6 \ ,
\label{libre}
\eeq
where  $E=M_{\rm ADM} - M_{\rm extremal}\ , M_{\rm extremal}=M_{\rm
ADM}(u_0=0)$.
Using that the M5-brane coordinate $x_5$ is
compactified on a circle with radius $R_0\lambda/N$, one has the relation
\beq
V_5={V_4 \lambda\over T_H N}\ .
\eeq
Expressing $u_0$ in terms of the string tension (\ref{sstt})
we obtain the intriguing relation
\beq
-{{\rm Action} \over V_4} = {1\over 12\pi } {N^2\over \lambda }
\sigma^2\ ,
\label{zzzz}
\eeq
that generalizes the result found in \cite{CORT98}
for the case of one angular momentum.
Thus, in terms of the string tension, the action
is independent of $a_{1,2}$.
It would be very interesting to have a derivation of (\ref{zzzz})
from the Yang--Mills side as a non-perturbative contribution to
the partition function (related to the expectation value of the
gluon condensate
$\langle {1\over 4g_{YM}^2} {\rm Tr\ }F_{\mu\nu}^2(0)\rangle $).

\subsection{The supersymmetric limit  $u_0 = 0$}

Metrics of rotating branes with non-extremality parameter $m=0$ greatly
simplify upon introducing Cartesian-type coordinates \cite{russo}.
For the extremal ($m=0$) M5-brane metric (2.1), one introduces \cite{KLT}
\beqa
&&y_1=\sqrt{r^2+l_1^2}\sin\theta \cos\varphi_1\ ,\ \ \  \ \ \ \ \ \ \ \
y_2=\sqrt{r^2+l_1^2}\sin\theta \sin\varphi_1\ ,\ \ \  \
\non\\
&&y_3=\sqrt{r^2+l_2^2}\cos\theta\sin\psi \cos\varphi_2\ ,\ \ \  \
y_4=\sqrt{r^2+l_2^2}  \cos\theta\sin\psi \sin\varphi_2\ ,\ \ \  \
\non\\
&& y_5=r \cos\theta\cos\psi\ .
\label{coor}
\eeqa
Using these coordinates we obtain
\beq
ds^2_{\rm IIA}=f^{-1/2}\big[-dx_0^2+\sum_{i=1}^4dx_i^2\big]
+f^{1/2}\sum_{j=1}^5 dy_j^2\ ,
\label{uuoo}
\eeq
where $f$ is obtained from Eq.~(\ref{zfff})
by taking the limit $\alpha\to \infty,\ m\to 0 $ at fixed $N$
using (\ref{sinhalpha}):
$f=1+{\pi N l_{\rm P}^3\over \Delta r^3}$,
 with $r$, $\theta ,\ \psi $
expressed in terms of $y_j$ by Eq.~(\ref{coor}).
In this limit the BPS bound is saturated,
$M_{\rm ADM}={\rm const}\  N$.
It can be shown that the function $f(y_j )$ satisfies the equation
$\partial_j \partial^j f=0$, i.e. it is a harmonic function in the 5-space
parameterized by $y_j$.
The metric (\ref{uuoo}) has unbroken supersymmetries, which can also be
understood
by interpreting the metric (\ref{uuoo}) as a multicenter distribution of BPS
D4-branes, by constructing the harmonic function $f$ as a linear
superposition of harmonic functions corresponding to each  D4-brane
\cite{KLT,sfetsos}.

The field theory limit of (\ref{uuoo}) can be obtained  by replacing
$f\to f-1$, and properly rescaling coordinates.
Alternatively, we can return to the metric (\ref{aaaa})
written in spherical coordinates, and set $u_0=0$.
The resulting metric has a curvature singularity in $u=a_1$
(we are assuming $a_1>a_2>0$), which cannot be removed by
any choice of periodicity in the $\tau $ coordinates
(the horizon region of the extremal $u_0=0$ metric
is not a Rindler space).
Because of the singularity, the supergravity approximation breaks down in
the $u_0=0$ case; in order to understand the
corresponding supersymmetric gauge theory, one needs to
understand the full string theory.
At the supergravity level, it is meaningless to associate a temperature to
this metric.

One can have control over the string-theory corrections if
we regularize the metric by taking $u_0\neq 0$ and
consider the limit of small $u_0$ (or equivalently,
$a_{1,2}/u_0$ large). For any value of $a_{1,2}/u_0$, one can choose
$\lambda $ sufficiently large so that all curvature invariants
are arbitrarily small. This is the technique used in the next section
when discussing the large $a_{1,2}/u_0$ limit.
Note that in this limit $T_H \to \infty $.
In this theory, all
fermions---which have masses ${\cal O}(T_H)$---decouple. On the other hand,
the spectrum of the $u_0=0$ theory 
must be supersymmetric
with the usual degeneracy between fermions and bosons.
The supersymmetric $u_0=0$ theory
cannot coincide with the theory obtained by taking the limit $u_0\to 0$ (in
the fashion described above), since the latter theory does not have fermions in
the spectrum.
Nevertheless, since the corresponding background metrics are
essentially the same (at $a_{1,2} \gg u_0$) it is possible that a part of the
structure of
the theory with  $a_1/u_0\gg 1$ may be  dictated
by the structure of the $u_0=0$ supersymmetric model.
We shall return to this point in our conclusions.

\section{Glueballs and the Related KK Modes}
\setcounter{equation}{0}
\setcounter{figure}{0}
\setcounter{table}{0}

\def\a{\alpha}
\def\b{\beta}
\def\uuu{\rho }
\def\del{\partial }
\def\qq{\qquad}
\def\m{\mu}
\def\n{\nu}
\def\th{\theta}
\newcommand{\eqn}[1]{(\ref{#1})}

The $0^{++}$ glueballs are related to spherically symmetric modes of
the dilaton fluctuations, of the form
\beq
\Psi = \phi(u) e^{ik\cdot x} \ ,
\label{oppg}
\eeq
where $M^2=-k^2$ \cite{WittenAdsThermal}.
The differential equation
determining the mass eigenvalues is obtained by substituting this into
the dilaton equation of motion
\beq
{1\ov \sqrt{g}} \del_\m \left[ e^{-2 \Phi} \sqrt{g} g^{\m\n} \del_\n \Psi
\right] = 0 \ ,
\label{dill}
\eeq
using the background metric (\ref{aaaa}),
and the formula
\beq
\sqrt{g}=C\ u^9 \Delta\cos^2\theta\sin\theta\sin\psi \ ,\ \ \ \ \
C={1\over 2\pi\lambda }
\left({4\pi\lambda A\over 3 u_0} \right)^6\ .
\eeq
In addition to the $0^{++}$ glueballs we consider particles
with non-vanishing $U(1)$ charge associated with the circle parameterized
by $\th_2$. The corresponding solutions of \eqn{dill} will be of the form
\beq
\Psi= \phi(u) e^{i k\cdot x} e^{i n \th_2}\ .
\label{kkth2}
\eeq
We will show both analytically (within the WKB approximation) and
numerically
that these states do decouple for a particular range of parameters.
We will also consider the KK states
associated with the $l=1$ modes of the $S^4$.
For the static ($a_1=a_2=0$) M5 metric, these
transform in the ${\bf 5}$ representation of $SO(5)$.
After introducing angular momentum, this decomposes
into ${\bf (2,1)}\oplus {\bf (1,2)} \oplus {\bf (1,1)}$ of
the Cartan subgroup $SO(2)\times SO(2)$.
According to this decomposition, the corresponding solutions
of \eqn{dill} for the two doublets will be given by
\beq
\Psi= \phi(u) e^{ik\cdot x} \sin\theta
\pmatrix{\cos\varphi_1\cr \sin\varphi_1}\ ,
\label{dob}
\eeq
\beq
\Psi= \phi(u) e^{ik\cdot x} \cos\theta \sin\psi
\pmatrix{\cos\varphi_2\cr \sin\varphi_2}\ ,
\label{dobb}
\eeq
whereas for the singlet it is of the form
\beq
\Psi=\phi(u) e^{ik\cdot x} \cos\theta \cos\psi \ .
\label{sing}
\eeq
In ordinary (finite $\lambda, N $) Yang--Mills theory there is no
$SO(2)\times SO(2)$
symmetry, so one would expect that at least the states
which transform non-trivially under $SO(2)\times SO(2)$
become very massive and decouple in the weak-coupling limit.
It is  clear that the singlet state (\ref{sing})
should also decouple. If it did not decouple at small $\lambda $,
it would then be represented by some (gluon field) operator in the gauge
theory.
In the zero angular momentum case,
this state combines with the other four components to form a multiplet
(a ${\bf 5}$)
of  $SO(5)$. Thus the singlet state cannot correspond to a purely gluonic
operator (since the gluon field is a singlet under  $SO(5)$),
and must decouple.

Finally, we shall also consider
$ 0^{-+}$ glueballs, which couple to the operator
$\tilde {\cal O}_4={\rm Tr}\ F\tilde F$.
On the D4-brane worldvolume,
the field that couples to this operator is the
R--R 1-form $A_\mu $, which satisfies the equation of motion
\beq
\del_\nu\big[ \sqrt{g} g^{\mu\rho}g^{\nu\sigma } (\del_\rho A_\sigma
- \del_\sigma A_\rho ) \big]
=0\ ,\qq \mu ,\nu =1,\dots,10\ .
\label{mmxx}
\eeq
Finding angular-independent solutions
is complicated, because of the non-diagonal components of the metric.
The metric becomes diagonal in the two opposite limits $a_{1,2}\ll u_0$
and $a_{1,2}\gg u_0$. In these cases one can consider
solutions of the form
\beq
A_{\rm \theta_2}=\chi_{\theta _2}(u)\ e^{ik\cdot x}\ ,\ \ \ \ \ \ \
A_\mu=0\ \ {\rm if}\ \
\mu\neq\theta_2\ .
\label{RReq}
\eeq

In the following we will first present the mass spectra of these states
obtained in the WKB approximation, and then the same spectra
obtained by using numerical methods. We present tables for each state
comparing the WKB with the numerical results and find that they are in a
very
good agreement. We also compare them to the lattice results
for the glueball states which were computed for $N=3$ and small $\lambda$.

\subsection{Mass spectrum in the WKB approximation}

In the following we use the WKB approach of \cite{RS}
(which generalizes the WKB approach of \cite{minahan})
to calculate the different mass spectra (including KK
modes) in the present case of QCD$_4$ with two angular momenta.
Consider differential equations of the form
\beq
\del_\uuu \left(f(\uuu ) \del_\uuu \phi\right) + \left(M^2 h(\uuu )
+ p(\uuu )\right) \phi = 0 \ ,
\label{diiff}
\eeq
where $M$ represents a mass parameter, and
$f(\uuu )$, $h(\uuu )$ and $p(\uuu )$, $\uuu\in [\uuu_H,\infty )$, are
three arbitrary functions
which are independent of $M$ and have the following behavior:
\beq
f\approx f_1 (\uuu -\uuu _H)^{s_1}\ ,\quad h\approx h_1 (\uuu -\uuu
_H)^{s_2}\ ,
\quad p\approx p_1 (\uuu -\uuu _H)^{s_3}\ , \qq {\rm as}\quad
\uuu\to \uuu_H\ ,
\label{lii0}
\eeq
\beq
f\approx f_2 \uuu^{r_1}\ ,\quad h\approx h_2 \uuu^{r_2}\ ,
\quad p\approx p_2 \uuu^{r_3}\ ,\qq {\rm as}\quad \uuu\to \infty\ ,
\label{lii1}
\eeq
where $r_{1,2,3}$, $s_{1,2,3}$, $f_{1,2}$, $h_{1,2}$
and $p_{1,2}$ are (real) numerical constants.
For large masses $M$, the
WKB method can be applied
to obtain the approximate spectrum. One finds \cite{RS}
\beq
M ^2= {\pi^2\ov \xi ^2}\ m \left(m + \Big(\!-\!1+ {\a_2\ov \a_1}
+ {\b_2\ov \b_1}\Big) \right)\ +\ {\cal O}(m^0) \ ,\qq m\geq 1\ .
\label{mxi}
\eeq
where
\beq
\xi=\int _{\uuu_H}^\infty d\uuu \ \sqrt{h\over f}\ ,
\eeq
is a constant which scales like a length, and
\beqa
&&\a_1= s_2-s_1 +2 \ , \qq \b_1= r_1-r_2-2 \ ,
\label{a1b1}\\
&&\a_2 = |s_1-1| \quad {\rm or} \quad \a_2=\sqrt{(s_1-1)^2 -4 {p_1\ov f_1} }
\
\
({\rm if}\ s_3-s_1+2=0)\ ,
\nonumber \\
&& \b_2 = |r_1-1| \quad {\rm or} \quad \b_2=\sqrt{(r_1-1)^2 -4 {p_2\ov f_2}}
\ \
({\rm if}\ r_1-r_3-2=0)\ .
\label{a2b2}
\eeqa
Consistency requires that $\a_1$ and $\b_1$ are strictly positive
numbers whereas
$s_3-s_1+2$ and $r_1-r_3-2$ can be either positive or zero.
Typically the validity of the WKB approximation requires that the quantum
number $m$ be much larger than 1 (for precise conditions see \cite{RS}).

\subsubsection{Masses of the  $0^{++}$ glueballs}

The masses of the $0^{++}$ glueballs are determined
from the differential equation \eqn{dill} with the ansatz \eqn{oppg}.
Introducing $\uuu =u^2$ one gets Eq. (\ref{diiff}) with\footnote{
In the rest of subsection 3.1
we will use the notation $a_{1,2}^2 =
b_{1,2}$.}

\beqa
&&f(\uuu)=(\uuu^2-b_1^2)(\uuu^2-b_2^2)-\uuu_0^3 \uuu \equiv
(\uuu-\uuu_H)(\uuu-\uuu_1)(\uuu-\uuu_2)(\uuu-\uuu_3)\ ,
\nonumber\\
&&h(\uuu)= {\uuu\over 4}\ ,\ \ \ \ p(\uuu)=0 \ .
\label{fhp4d}
\eeqa
This gives for the various constants
\beqa
&&s_1 =1 \ ,\quad s_2=0\ ,\quad r_1=4 \ ,\quad r_2=1 \ ,
\nonumber\\
&&\alpha_1=1\ ,\quad \alpha_2=0 \ ,\quad \beta _1=1\ ,\quad \beta _2=3\ .
\eeqa
Using (\ref{mxi}) one obtains
the following mass spectrum
\beqa
&& M^2= {\pi ^2\over \xi ^2}\ m(m+2)+ {\cal O}(m^0)\ ,\qq m\geq 1\ ,
\nonumber \\
&& \xi = {1\over 2} \int _{\uuu_H}^\infty {d\uuu \ \sqrt{\uuu}\over
\sqrt{(\uuu-\uuu_H)(\uuu-\uuu_1)(\uuu-\uuu_2)(\uuu-\uuu_3)} }\ ,
\label{sadj}
\eeqa
This formula implies that mass ratios between resonances
are, in the WKB approximation, independent of the angular momentum
parameters $b_1, b_2$. This is similar to ${\rm QCD}_3$, where the general
rotating D3-brane solution with three angular momenta parameters was used
\cite{RS}. As in \cite{RS}, the WKB approximation breaks down in the region
near $b_1=b_2$, $u_0=0$.

\subsubsection{The KK modes on the circle}

For the KK modes with non-vanishing $U(1)$ charge corresponding
to
the periodic variable $\th_2$ we look for solutions of \eqn{dill} with the
ansatz \eqn{kkth2}. In this case we obtain Eq. \eqn{diiff} with $M^2$
replaced by $M^2-4 \pi^2 n^2 T_H^2$ and
\beqa
&&f(\uuu)=(\uuu^2-b_1^2)(\uuu^2-b_2^2)-\uuu_0^3 \uuu \ ,
\nonumber\\
&&h(\uuu)= {\uuu\over 4}\ ,
\qq p(\uuu)= -{\pi^2 n^2 \uuu_0^3 T_H^2 \uuu^2\ov f(\uuu)} \ .
\label{fhp4d1}
\eeqa
This gives for the various constants
\beqa
&&s_1 =1 \ ,\quad s_2=0\ ,\quad s_3=-1\ ,\quad r_1=4 \ ,\quad r_2=1 \ ,
\quad r_3=-2 \ ,
\nonumber\\
&&\alpha_1=1\ ,\quad \alpha_2= {2\pi n \uuu_0^{3/2} \uuu_H T_H\ov
(\uuu_H-\uuu_1)(\uuu_H-\uuu_2)(\uuu_H-\uuu_3)} \ ,\quad
\beta _1=1\ ,\quad \beta _2=3\ .
\label{dhf}
\eeqa
Using \eqn{teeem} and \eqn{aaa1}
we see that $\alpha_2=n$. Then
(\ref{mxi}) (with $M^2\to M^2-4\pi^2 n^2 T_H^2$) gives
the following mass spectrum
\beq
M^2 = 4\pi^2 n^2 T_H^2 \ + \
{\pi^2\over \xi^2 }\ m(m+2+ n)+ {\cal O}(m^0)\ ,\qq m\geq 1\ .
\nonumber \\
\label{sadj1}
\eeq
We would like to examine the way that these states decouple in two limiting
cases. First consider the case with $b_1\gg \uuu_0\ {\rm and}\ b_2$.
Then we see from \eqn{teeem} that
\beq
T_H\simeq { b_1^2\ov \pi \uuu_0^{3/2}}~.
\eeq
On the other hand, in the same limit,
\beq
\xi \simeq {1.31 \ov b_1^{1/2}}~.
\eeq
Therefore the ratios of the masses of the glueballs to those of the $U(1)$
charged particles behave as
\beq
{M_{\rm glueb.}\ov M_{\rm circ.}} \simeq 1.20 {\sqrt{m(m+2)}\ov n}
\left({\uuu_0\ov b_1}\right)^{3/2}\ ,\qq {\rm for}\quad b_1\gg \uuu_0\
{\rm and }\ b_2\ .
\label{deed}
\eeq
Hence, the KK modes on the circle decouple with a power law.
Now consider the case $b_1=4 b_2 \gg \uuu_0$.
Then
\beq
T_H\simeq { 15 b_1^2\ov 16 \pi \uuu_0^{3/2}}
\eeq
 and
\beq
\xi \simeq {1.33\ov b_1^{1/2}}~.
\eeq
Therefore the ratios of the masses of the glueballs to those of the $U(1)$
charged particles behave as
\beq
{M_{\rm glueb.}\ov M_{\rm circ.}} \simeq 1.26 {\sqrt{m(m+2)}\ov n}
\left({\uuu_0\ov b_1}\right)^{3/2}\ ,\qq {\rm for}\quad
b_1=4 b_2 \gg \uuu_0\ ,
\label{deed1}
\eeq
showing that in this case there is also decoupling with the same power
law as in \eqn{deed} (up to a slightly different numerical factor).

\subsubsection{The KK modes of $S^4$}

Let us now compute the mass spectrum for the KK states
with non-trivial angular dependence on $S^4$.
The two equations corresponding
to the doublets must be related
by an interchange of $a_1$ and $a_2$, whereas the one corresponding to the
singlet should be invariant under such an interchange.
For the $({\bf 2},{\bf 1})$ doublet  we make the ansatz \eqn{dob}.
Inserting into Eq. \eqn{dill}
one finds equation (\ref{diiff})
with
\beqa
&&f(\uuu)=(\uuu^2-b_1^2)(\uuu^2-b_2^2)-\uuu_0^3 \uuu\ ,
\qq h(\uuu)\ = \ {\uuu\over 4}\ ,
\nonumber\\
&&p(\uuu)= -4 \uuu^2 \left(1-{b_2^2\ov 2 \uuu^2}\right)
- {b_1^2 \uuu^4\ov f(\uuu)}\ \left(1-{b_2^2\ov \uuu^2}\right)^2\ .
\label{fhps1}
\eeqa
For the various constants we find
\beqa
&& s_1=1 \ ,\quad s_2 =0 \ , \quad s_3 =- 1\ , \quad r_1 =4 \ ,\quad
r_2 =1 \ ,\quad r_3 =2 \ ,
\nonumber \\
&& f_1=(\uuu_H-\uuu_1)(\uuu_H-\uuu_2)(\uuu_H-\uuu_3) \ ,\quad f_2 =1 \ ,
\quad p_1 =
-{b_1^2 \uuu_H^4\ov f_1} \left(1-{b_2^2\ov \uuu_H^2}\right)^2\ ,
\nonumber \\
&& p_2=-4\ ,\quad \alpha_1=1 \ ,\quad \alpha_2 = 2 {b_1 \uuu_H^2\ov f_1}
\left(1-{b_2^2\ov \uuu_H^2}\right)\ ,
\quad \b_1=1\ , \quad \b_2 =5 \ .
\label{sajh12}
\eeqa
Hence, the mass is given by
\beq
M^2= {\pi ^2\ov \xi ^2}\ m\left(m+4 +
{2b_1 \uuu_H^2\ov f_1} \Big(1-{b_2^2\ov \uuu_H^2}\Big)\right) \
+\ {\cal O}(m^0)\ ,\qq m\ge 1\ ,
\label{hsj12}
\eeq
where $\xi $ is given by \eqn{sadj}.
Consider the mass formula in the
region $b_1\gg \uuu_0$
where the KK modes on the circle decouple, as
in the case of one angular momentum \cite{russo,CORT98}.
Using
$\uuu _H\cong b_1$, the mass formula takes the form
\beq
M^2 \simeq {\pi ^2\ov \xi ^2}\ m(m+ 5)\ + \ {\cal O}(m^0)\ ,\qq m\ge 1\ .
\label{sqs3}
\eeq
This shows that for $b_1\gg \uuu_0$ the mass of these KK
states is of the same order as the glueball masses \eqn{sadj}.

For the $({\bf 1},{\bf 2})$ doublet  we make the ansatz \eqn{dobb}
We obtain the same results as \eqn{fhps1}--\eqn{hsj12} with $b_1$ and
$b_2$ interchanged. For completeness
we include the mass formula
\beq
M^2= {\pi ^2\ov \xi^2}\ m\left(m+4 +
{2b_2 \uuu_H^2\ov f_1} \Big(1-{b_1^2\ov \uuu_H^2}\Big)\right) \
+\ {\cal O}(m^0)\ ,\qq m\ge 1\ .
\label{hs13}
\eeq
For $\uuu_H\cong b_1$ this becomes
\beq
M^2\simeq {\pi ^2\ov \xi ^2}\ m(m+ 4)\ + \ {\cal O}(m^0)\ ,\qq m\ge 1\ ,
\label{sqs4}
\eeq
where $\xi$ is given by \eqn{sadj}.
This shows that for $b_1\gg \uuu_0$ the mass of these KK
states is of the same order as glueball masses and a little lighter than the
modes corresponding to the KK doublet \eqn{dob}.
As a general rule (which applies in particular to QCD$_3$ \cite{RS}),
states with $\varphi$-dependence corresponding
to the largest angular parameter are slightly heavier.
In addition to the two doublets  there is also a
singlet $({\bf 1},{\bf 1})$, represented by Eq.~\eqn{sing}.
We find that the function $\phi(\uuu)$ obeys \eqn{diiff} with
\beqa
&&f(u)=(\uuu^2-b_1^2)(\uuu^2-b_2^2)-\uuu_0^3 \uuu\ ,
\qq h(\uuu) = {\uuu\over 4}\ ,
\nonumber\\
&&p(\uuu)= 2(b_1^2+b_2^2-2 \uuu^2)\ .
\label{fhps2}
\eeqa
For the various constants necessary to compute the corresponding masses we
find
\beqa
&& s_1=1 \ ,\quad s_2 =0 \ , \quad s_3 =0\ , \quad r_1 =4 \ ,\quad
r_2 =1 \ ,\quad r_3 =2 \ ,
\nonumber \\
&& p_2=-4\ ,\quad f_2=1\ ,\quad \alpha_1=1 \ ,\quad \alpha_2 = 0\ ,
\quad \b_1=1\ , \quad \b_2 =5 \ .
\label{sajh14}
\eeqa
Using \eqn{mxi} the mass formula for the singlet~\eqn{sing} reads
\beq
M^2= {\pi ^2\over \xi^2}\ m(m+ 4)\ + \ {\cal O}(m^0)\ ,\qq m\ge 1\ ,
\label{sq23}
\eeq
where $\xi$ is again given in \eqn{sadj}. Clearly, the masses of
these modes are of the same order as the glueball masses \eqn{sadj},
albeit slightly heavier.

\subsubsection{Masses of the $0^{-+}$ glueballs}

Let us finally also consider
$ 0^{-+}$ glueballs. As we have mentioned, finding angular-independent
solutions is complicated, because of the non-diagonal components of the
metric.
The metric becomes diagonal in the two opposite limits $b_{1,2}\ll \uuu_0$
and $b_{1,2}\gg \uuu_0$. In these cases one can consider
solutions of the form \eqn{RReq}.
Substituting this into \eqn{mmxx},
we obtain a second order ordinary differential equation
which, upon introducing $\rho=u^2$ and writing $a^2_{1,2}= b_{1,2}$,
can be written as Eq. \eqn{diiff} (with $\phi(\uuu)\to \chi_{\th_2}(\uuu)$~)
with
\beqa
&& f(\uuu) = (\uuu^2-b_1^2)(\uuu^2-b_2^2)\ ,\qq p(\uuu)=0\ ,
\nonumber \\
&& h(\uuu) = {1\ov 4}
{\uuu (\uuu^2-b_1^2)(\uuu^2-b_2^2)\ov (\uuu^2-b_1^2)(\uuu^2-b_2^2)-
\uuu_0^3 \uuu}\ .
\label{qre}
\eeqa
This gives for the various constants
\beqa
&&s_1 =0 \ ,\quad s_2=-1\ ,\quad r_1=4 \ ,\quad r_2=1 \ ,
\nonumber\\
&&\alpha_1=1\ ,\quad \alpha_2=1 \ ,\quad \beta _1=1\ ,\quad \beta_2=3\ .
\eeqa
Using (\ref{mxi}) one obtains
the following mass spectrum
\beq
M^2 = {\pi^2\over \xi^2 } m(m+3)+ {\cal O}(m^0)\ ,\qq m\geq 1\ ,
\nonumber \\
\label{sadj5}
\eeq
where, it turns out,
the constant $\xi$ is still given by the corresponding expression in
\eqn{sadj}. In the limit when $b_1,b_2\gg \rho_0$, the singularity structure
of Eq.~(\ref{diiff}) changes \cite{minahan}.
In this limit however, the equation
coincides with the equation for the $0^{++}$ glueballs, and therefore
the mass formula should be changed to
\beq
M^2 = {\pi^2\over \xi^2 } (m+1)(m+3)+ {\cal O}(m^0)\ ,\qq m\geq 1\ ,
\nonumber \\
\label{sadj6}
\eeq
corresponding to the mass formula (\ref{sadj})
of the $0^{++}$ glueballs shifted by
one (since the lowest state should correspond to the zero mode of
Eq.~(\ref{diiff}) and not to a glueball state \cite{Wittentheta}).
This will be the
formula used for comparison to the numerical results for $a\gg u_0$.


\subsection{Numerical evaluation of the mass spectra}
\setcounter{figure}{0}
\setcounter{table}{0}

In the following, we present the results of the numerical evaluation of the
mass spectra corresponding to the states described in Section 3.1. For every
state,
we will illustrate the dependence of the masses on the angular momentum
parameter along a generic direction (chosen to be $a_1=2a_2$), along
the special direction $a_1=a_2$, and a table comparing the numerical and WKB
results (and the lattice results for glueball states).

\subsubsection{Masses of the $0^{++}$ glueballs}

The equation for the $0^{++}$ glueballs can also be written as
\beqa
\partial_u \left[ \frac{1}{u}\left( (u^4-a_1^4)(u^4-a_2^4)-u^2 u_0^6\right)
\partial_u f(u)\right] -k^2u^3f(u)=0.
\label{ggbb}
\eeqa
This equation is symmetric under the interchange of $a_1$ and $a_2$, and
reproduces Eq.~(2.14) of Ref.~\cite{CORT98} for $a_2\to 0$.
This differential equation can be solved numerically using the shooting
method as described in Ref.~\cite{COOT98}.
We require that the solution be normalizable (that is
for $u\to \infty$ $f(u)$ should vanish), and regular at the horizon $u_H$.
These conditions restrict the possible values of $M^2$ to a discrete set,
which can be identified with the glueball masses. The analysis
of~\cite{CORT98}
demonstrated that the $0^{++}$ glueball masses are very stable against the
variation of a single angular momentum parameter. The numerical solutions of
Eq. (\ref{ggbb}) show that this statement remains valid for the whole
range of angular parameters $(a_1,a_2)$, except in the region
$a_1=a_2\gg u_0$.
This is consistent with the fact that in the WKB approximation
(to order $1/m$)
the ratio of masses are independent of $a_1,a_2$ everywhere except at
$a_1=a_2\gg u_0$, where the approximation breaks down.
As mentioned before, $a_1=a_2$, $u_0=0$ is the special
region
where the inner horizon coincides with the outer horizon.
As a result, the factor multiplying the second derivative term in
(\ref{ggbb}) has a double zero (instead of simple zero), and the behavior of
the solutions is different.

\begin{figure}
\PSbox{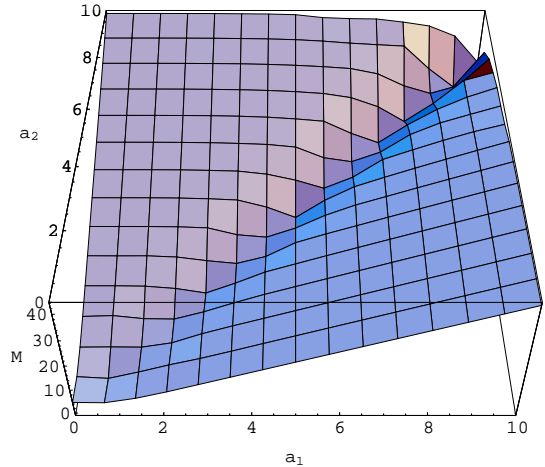 hscale=70 vscale=70 hoffset=100
voffset=-15}{15cm}{5cm}
\caption{The unnormalized values of the $0^{++}$ glueball mass (the lowest
eigenvalues of Eq. (\protect\ref{ggbb})) as a
function of the two angular momenta.
Note that this function is smooth everywhere
except in the region $a_1=a_2\rightarrow\infty$.
\label{fig:gbmass}}
\end{figure}

\begin{figure}
\PSbox{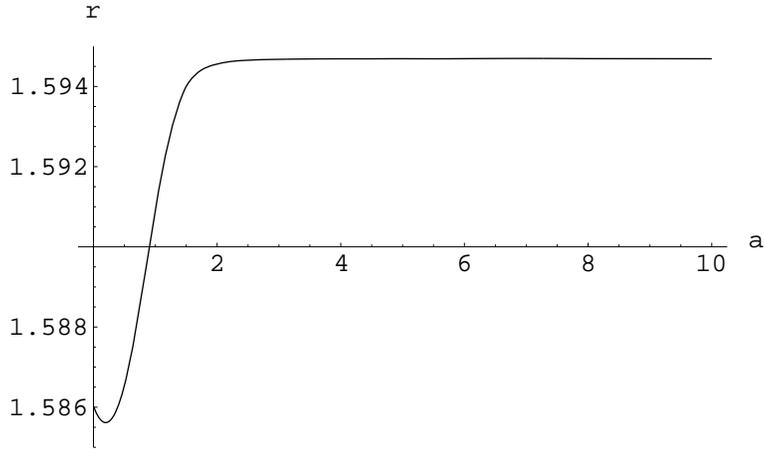 hscale=100 vscale=100 hoffset=50
voffset=-60}{13.7cm}{5.5cm}
\caption{The ratio of the $0^{++*}$ mass to the $0^{++}$ mass along a
generic
direction, chosen here to be $a_1=2a_2=a$.
Note, that the change in the ratio is tiny, and the asymptotic value of the
ratio is the same as in Ref.~\protect\cite{CORT98} in the case of
$a_1\rightarrow
\infty$,
$a_2=0$.
\label{fig:noneq}}
\end{figure}

In Fig.~\ref{fig:gbmass} we show the
behavior of the lowest eigenvalue of Eq. (\ref{ggbb}).
The valley along $a_1=a_2$ is related to the fact that
the differential equation (and the physics of the model)
is symmetric under the interchange $a_1\leftrightarrow a_2$.
Note that the function
is smooth except at the point $a_1=a_2=\infty $
(or $a_1=a_2$, $u_0=0$). In Fig.~\ref{fig:noneq} we
show the behavior of the ratio of the glueball masses along the direction
$a_1=2a_2$, which illustrates the fact that along a generic direction
(by a generic direction we mean that it does not asymptote to
$a_1=a_2$)
the glueball mass ratios behave just like for the case with only one angular
momentum, that is they change only slightly and take on their asymptotic
value
very quickly. The fact that the behavior of the glueball masses
does not change can be seen comparing Fig.~\ref{fig:noneq} to Fig. (2.1)
of Ref.~\cite{CORT98}. Table~\ref{tab:opp} contains the comparison
of the lattice results, the numerical solutions and the WKB results
along a generic direction (chosen to be $a_1=2a_2$).

\begin{figure}
\PSbox{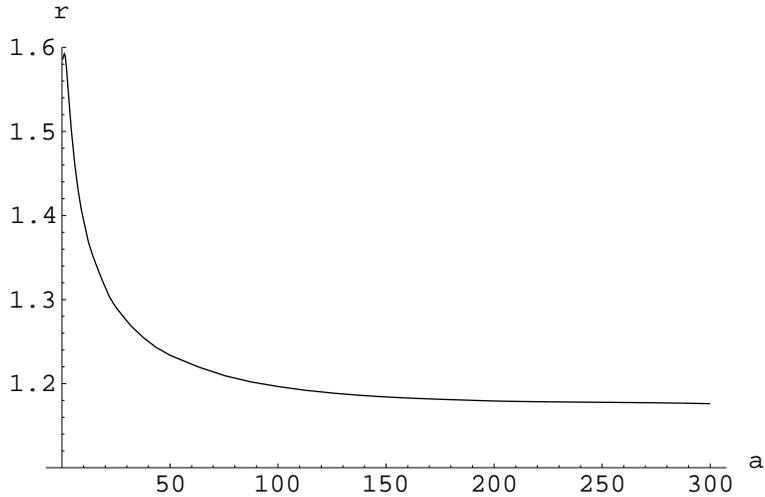 hscale=100 vscale=100 hoffset=50
voffset=-60}{13.7cm}{6cm}
\caption{The behavior of the ratio $r$
of the mass of the excited $0^{++*}$ glueball mass to
the $0^{++}$ mass along the line
$a_1=a_2$.
Note, that along this direction the solutions behave very differently
than anywhere else in the parameter space and depart significantly from
the lattice results.
\label{fig:equalang}}
\end{figure}

In Fig.~\ref{fig:equalang} we show a ratio of masses along the special
direction $a_1=a_2$, where the inner and outer horizons come together
as $a_1/u_0\to\infty $ (this is also the region where the WKB approximation
breaks down). In this region
the mass ratios behave very differently than anywhere else and
depart from the lattice results
(for example, the weak-coupling lattice value for $M_{0^{++*}}/M_{0^{++}}$
for $N=3$ is about $1.74$, which is notably bigger than the numbers
of Fig.~\ref{fig:equalang} at large $a$).

\begin{table}
\[
\begin{array}{|l|c|c|c|}
\hline
\mbox{state} & \mbox{lattice} & \mbox{numerical} & WKB \\ \hline
0^{++} & 1.61 \pm 0.15 & 1.61\; \mbox{(input)}& 1.55 \\ \hline
0^{++*}& 2.8 & 2.57 & 2.53\\ \hline
0^{++**}& - & 3.49 & 3.46\\ \hline
0^{++***}& - & 4.40 & 4.37\\ \hline
0^{++****}& - & 5.30 & 5.28\\ \hline
0^{++*****}& - & 6.20 & 6.18 \\ \hline
\end{array}
\]
\caption{The masses of the first few $0^{++}$ glueballs in GeV. The first
column gives the available lattice results
\protect\cite{Teper97,MorningstarPeardon,Peardon}, the second the
asymptotic value of the supergravity calculation using the numerical method
(the point is chosen to be $a_1=2a_2=20u_0$), while the third column the
WKB result for the same supergravity approximation.\label{tab:opp}}
\end{table}

\bigskip

\subsubsection{Masses of the $0^{-+}$ glueballs}

\begin{figure}
\PSbox{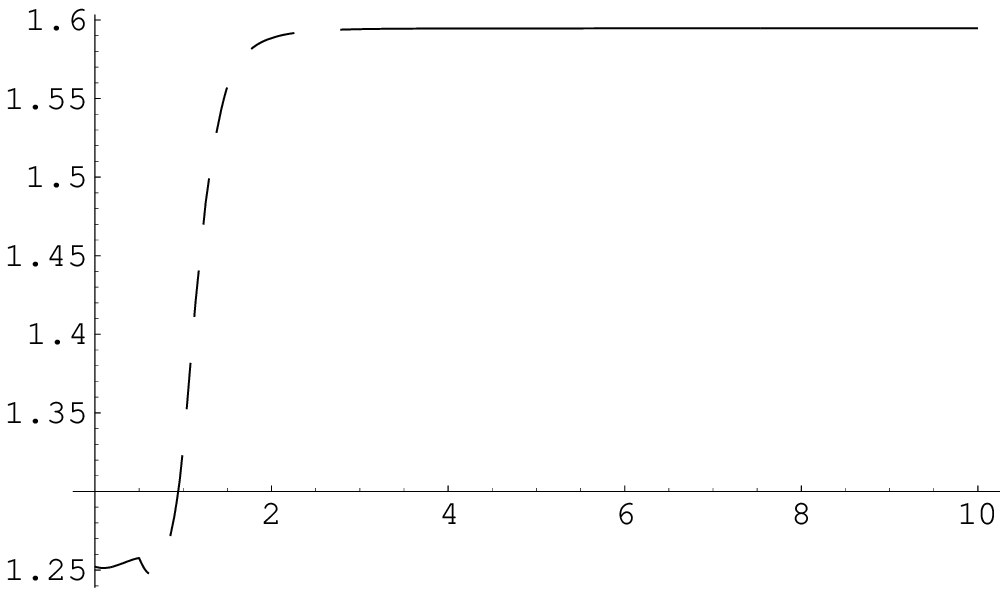 hscale=100 vscale=100 hoffset=50
voffset=-60}{13.7cm}{5.5cm}
\caption{The ratio of the lowest $0^{-+}$ mass to the lowest
$0^{++}$ mass along a
generic direction, chosen here to be $a_1=2a_2=a$.
Note, that the ratio is very stable against the variations of the
parameters. The actual change in the ratio is sizeable, and independent
of the direction chosen in the $(a_1,a_2)$ parameter space (except the
line $a_1=a_2$) and agrees with the ratio found in
Ref.~\protect\cite{CORT98}
for the case of $a_1 \rightarrow \infty$, $a_2=0$. As explained in the text,
this figure is only reliable for the regions $a \ll u_0$ and $a \gg u_0$ which
are shown by solid lines, while for the intermediate region denoted by a
dashed line there are corrections due to the non-vanishing off-diagonal
components of the metric.
\label{fig:ompgen}}
\end{figure}

The differential equation for $0^{-+}$
glueballs can be written as
\beq
{1\ov u^3} \del_u
\big[{1\over u} (u^4-a^4_1)(u^4-a^4_2) \del_u\chi_{\theta_2}(u)\big]
=-M^2 {(u^4-a^4_1) (u^4-a^4_2) \over (u^4-a^4_1)
(u^4-a^4_2)-u_0^6u^2 } \ \chi_{\theta_2}(u)\ .
\label{0-+}
\eeq
The corresponding mass spectrum can be obtained
using a similar
numerical method as for the $0^{++}$ glueballs. The dependence of
the lightest $0^{-+}$ glueball mass on the angular momentum along a generic
direction (chosen again to be $a_1=2a_2$) is given in Fig.~\ref{fig:ompgen}.
One can see that while the masses are fairly stable against variations
of the angular momentum, just like in the case of $a_2=0$ discussed in
\cite{CORT98}, the actual values
of the mass ratios compared to $0^{++}$ increase by a sizeable ($\sim 25\%$)
value. The change is in the right direction as suggested by recent
improved lattice simulations \cite{Peardon}.
The actual asymptotic value of the mass ratio
$\frac{m_{0^{-+}}}{m_{0^{++}}} =1.59$ is the same as for the $a_2=0$ case
everywhere except very close to the region $a_1=a_2 \gg u_0$.
Table~\ref{tab:omp} contains the comparison of the lattice results to the
supergravity results evaluated using the numerical and the WKB methods.

\begin{figure}
\PSbox{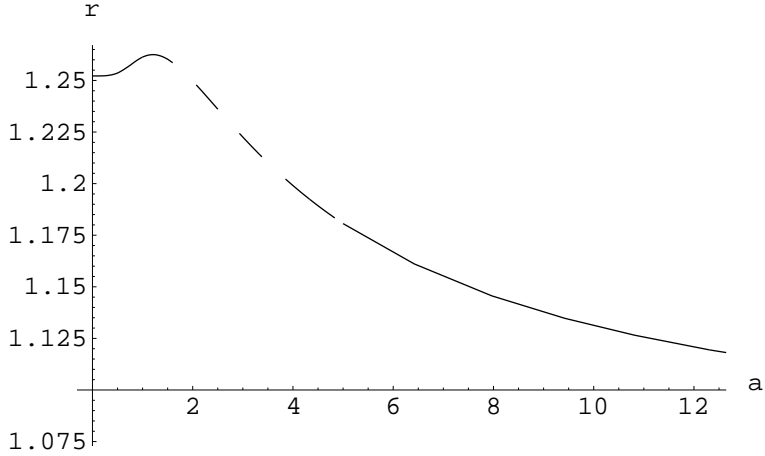 hscale=100 vscale=100 hoffset=50
voffset=-60}{13.7cm}{4.5cm}
\caption{The behavior of the ratio $r$
of the mass of the lowest $0^{-+}$ glueball mass to
the lowest $0^{++}$ mass along the line
$a_1=a_2$. As explained in the text,
this figure is only reliable for the
regions $a \ll u_0$ and $a \gg u_0$ which
are shown by solid lines.
\label{fig:ompspec}}
\end{figure}

Just as
for the case of the $0^{++}$ glueballs, this ratio behaves very differently
along the special $a_1=a_2$ direction, and departs significantly from the
lattice result $(\frac{m_{0^{-+}}}{m_{0^{++}}})_{lattice} =1.46$ as it can
be seen in Fig.~\ref{fig:ompspec}.

\begin{table}
\[
\begin{array}{|l|c|c|c|c|}
\hline
\mbox{state} & \mbox{lattice} & \mbox{Num.} (a_{1,2}=0) &
\mbox{Num.} (a_1=2a_2=20 )
& \mbox{WKB} \\ \hline
0^{-+} & 2.59 \pm 0.13 & 2.00 & 2.57 &  2.53\\ \hline
0^{-+*} & 3.64 \pm 0.18 & 2.98 &3.49 & 3.46\\ \hline
0^{-+**} & - & 3.91 & 4.40 & 4.37 \\ \hline
0^{-+***} & - & 4.83 & 5.30 & 5.28 \\ \hline
0^{-+****} & - & 5.74 & 6.20 & 6.18 \\ \hline
0^{-+*****}& - & 6.64 & 7.10 & 7.09  \\ \hline
\end{array}
\]
\caption{The masses of the first few $0^{-+}$ glueballs in GeV.
Unlike $0^{++}$ glueballs, the supergravity masses for these glueballs
are sensitive to the values of $a_1, a_2$.
Two cases are displayed, illustrating the typical values that one gets for
small and large $a_1, a_2$ (the asymptotic values for large $a_1,a_2$ are
the same for any generic direction, i.e. with $a_1\neq a_2$).
The lattice results are from
\protect\cite{MorningstarPeardon,Peardon}.
\label{tab:omp}}
\end{table}

\subsubsection{Masses of the KK modes of $S^4$}

\begin{figure}
\PSbox{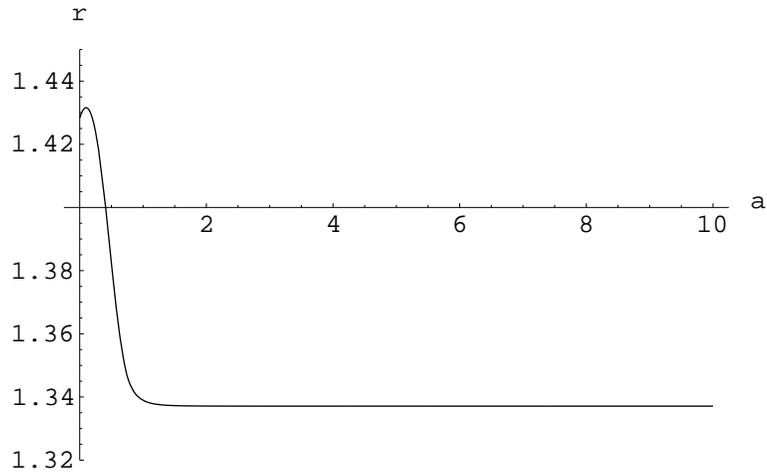 hscale=100 vscale=100 hoffset=50
voffset=-60}{13.7cm}{6.5cm}
\caption{The mass ratio $r$ of the $SO(2)\times SO(2)$
singlet KK mode to the lowest $0^{++}$ glueball along the
generic direction $a_1=2 a_2$.
\label{fig:singletgen}}
\end{figure}

\begin{figure}
\PSbox{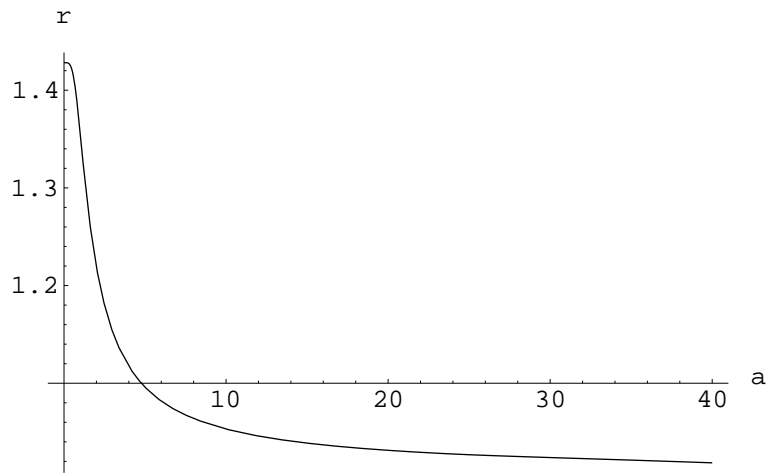 hscale=100 vscale=100 hoffset=50
voffset=-60}{13.7cm}{6.5cm}
\caption{The mass ratio $r$ of the $SO(2)\times SO(2)$
singlet KK mode to the lowest glueball mass $0^{++}$
along the special direction $a_1=a_2$.
\label{fig:singletspec}}
\end{figure}

\begin{figure}
\PSbox{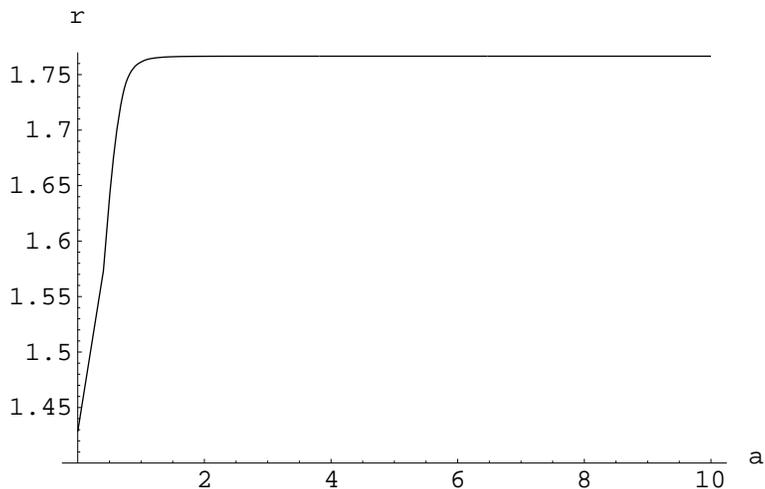 hscale=100 vscale=100 hoffset=50
voffset=-60}{13.7cm}{6.5cm}
\caption{The mass ratio $r$ of the $SO(2)\times SO(2)$ doublet KK mode
to
the lowest glueball masses
along the generic direction $a_1=2 a_2$.
\label{fig:kk4dgen}}
\end{figure}

\begin{figure}
\PSbox{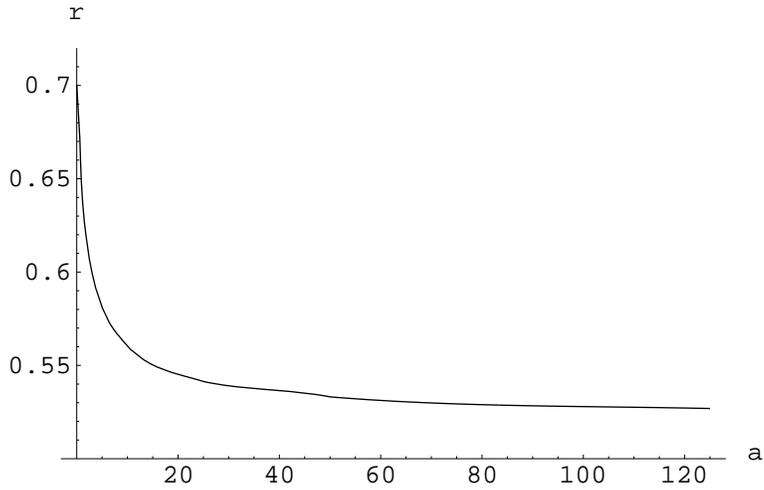 hscale=100 vscale=100 hoffset=50
voffset=-60}{13.7cm}{6.5cm}
\caption{The mass ratio $r$ of the lightest glueball to the
$SO(2)\times SO(2)$ doublet KK modes
along the special direction $a_1=a_2$.
\label{fig:kk4dspec}}
\end{figure}

In terms of $u=\sqrt{\rho} $,
equation \eqn{fhps1} for the first KK
doublet (\ref{dob}) reads
\beq
\partial_u \left[ \frac{1}{u}\left( (u^4-a_1^4)(u^4-a_2^4)-u^2u_0^6\right)
\partial_u f(u)\right] -u^3 \left( k^2+ {4 u^2\over h_0} H(u) \right) f(u)
=0.
\label{ttz}
\eeq
where
\beq
h_0(u)= \big(1-{a_1^4\over u^4}\big) \big( 1-{ a_2^4\over u ^4}
\big)-{u_0^6\over u^6}\ ,
\eeq
\beq
H(u)=4 h_0(u)\ \big( 1- {a_2^4\over 2 u^4} \big)+ {a_1^4\over u^4}
\big(1- {a_2^4\over u^4}\big)^2\ .
\eeq
The components of the second doublet (\ref{dobb}) give the same equation
with $a_1$ and $a_2$ interchanged.
Finally, the equation that determines the mass spectrum of the
singlet (\ref{sing}) is
\beqa
u^2 \partial_u \left[ \frac{1}{u}\left( (u^4-a_1^4)(u^4-a_2^4)-u^2\right)
f'(u)\right] + u^3 (8 a_1^4 + 8 a_2^4 - k^2 u^2 - 16 u^4) f(u) =0.
\label{tyy}
\eeqa
This is symmetric under the interchange of $a_1$ and $a_2$.
One can again numerically determine the solutions of these equations using
the
shooting method. In Figs. \ref{fig:singletgen} and \ref{fig:singletspec}
we show the behavior of the
$SO(2)\times SO(2)$ singlet mode first along a generic direction (which was
again chosen to be $a_1=2a_2$), and then along the special direction
$a_1=a_2$. One can see that this mode does not decouple on any region of the
parameter space. Figs. \ref{fig:kk4dgen} and \ref{fig:kk4dspec} show the
similar plots for the non-singlet KK modes (for \eqn{ttz}),
which similarly do not decouple
anywhere in the parameter space. Tables \ref{tab:kksing} and \ref{tab:kknon}
show the comparison of the first few KK modes evaluated using the numerical
and the WKB methods.

\begin{table}
\[
\begin{array}{|l|c|c|}
\hline
\mbox{state} & \mbox{numerical} & WKB \\ \hline
KK & 2.15 & 2.00 \\ \hline
KK^{*}& 3.23 & 3.09\\ \hline
KK^{**}& 4.20 & 4.09\\ \hline
KK^{***}& 5.15 & 5.05\\ \hline
KK^{****} & 6.07 & 5.99\\ \hline
KK^{*****}& 6.99 & 6.91\\ \hline
\end{array}
\]
\caption{The masses of the first few singlet KK
modes in GeV. The first
column gives the asymptotic value of the supergravity calculation using the
numerical method
(the point is chosen to be $a_1=2a_2=20u_0$), while the second column the
WKB result for the same supergravity approximation.\label{tab:kksing}}
\end{table}

\begin{table}
\[
\begin{array}{|l|c|c|}
\hline
\mbox{state} & \mbox{numerical} & WKB \\ \hline
KK & 2.84 & 2.19 \\ \hline
KK^{*}& 3.80 & 3.34\\ \hline
KK^{**}& 4.73 & 4.37\\ \hline
KK^{***}& 5.54 & 5.36\\ \hline
KK^{****}& 6.57 & 6.31\\ \hline
KK^{*****} & 7.47 & 7.25\\ \hline
\end{array}
\]
\caption{The masses of the first few non-singlet KK
modes in GeV. The first
column gives the asymptotic value of the supergravity calculation using the
numerical method
(the point is chosen to be $a_1=2a_2=20u_0$), while the second column the
WKB result for the same supergravity approximation.\label{tab:kknon}}
\end{table}

\subsubsection{Masses of the KK modes on the circle}

\begin{figure}
\PSbox{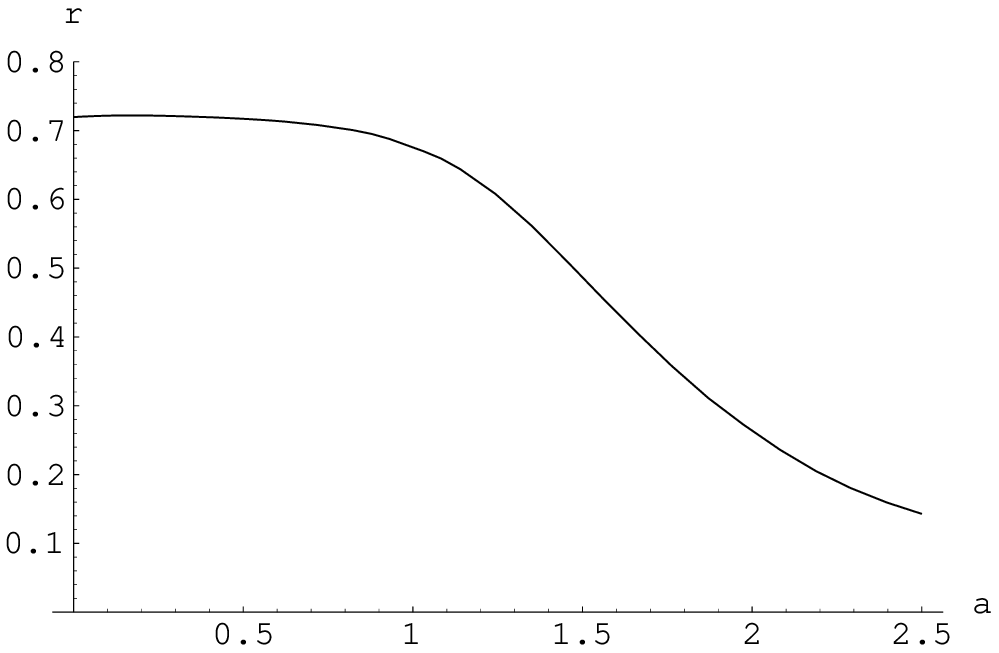 hscale=100 vscale=100 hoffset=50
voffset=-60}{13.7cm}{6.5cm}
\caption{The mass ratio $r$ of the lightest glueball to the KK mode on
the compact D-brane coordinate $\theta_2$
along the generic direction $a_1=2 a_2=a$.
Just as for the case with only one angular momentum, these states decouple
very quickly from the spectrum.
\label{fig:kkcircnon}}
\end{figure}

\begin{figure}
\PSbox{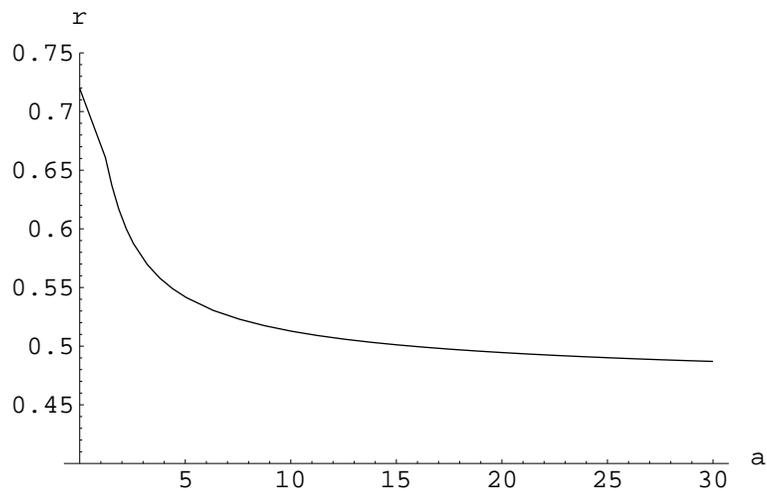 hscale=100 vscale=100 hoffset=50
voffset=-60}{13.7cm}{5.5cm}
\caption{The mass ratio $r$ of the
lightest glueball to the KK mode on
the compact D-brane coordinate $\theta_2$
along the special direction $a_1=a_2$.
\label{fig:kkcirc}}
\end{figure}

Next we consider the KK modes coming from the compact D-brane
coordinate. These modes have the form (\ref{kkth2}),
where $\phi (u )$ obeys the differential equation
\begin{eqnarray}
\partial_u \left[ \frac{1}{u}\left( (u^4-a_1^4)(u^4-a_2^4)-u^2\right)
\partial_u \phi(u)\right] =\phi (u)u^3\bigg( k^2+\frac{9
n^2(a_1^4-u^4)(a_2^4-u^4)}{A^2(u^8-
u^4(a_1^4+a_2^4)-u^2+a_1^4a_2^4)} \bigg)\ .
\hspace*{-1cm}\nonumber \\
\end{eqnarray}
One can again numerically solve these equations. For a generic direction
(chosen to be again $a_1=2a_2$) we find that these modes decouple very
quickly
from the spectrum, just like in the case with one angular momentum
parameter
discussed in Ref.~\cite{CORT98}. This is illustrated in
Fig.~\ref{fig:kkcircnon}. For the case of the special direction $a_1=a_2$,
the numerical analysis of the decoupling is inconclusive. The masses of
these
KK modes grow much slower than for the generic case. At the point when
our numerical solutions become unreliable, these modes are not decoupled
yet, however one can not rule out the possibility that for $a\to \infty$
they
eventually do decouple (see Fig.~\ref{fig:kkcirc}).

\begin{table}
\[
\begin{array}{|l|c|c|}
\hline
\mbox{state} & \mbox{numerical} & WKB \\ \hline
KK & 11.27 & 11.24 \\ \hline
KK^{*}& 11.48 & 11.45\\ \hline
KK^{**}& 11.76 & 11.72\\ \hline
KK^{***}& 12.09 & 12.06\\ \hline
KK^{****}& 12.48 & 12.45\\ \hline
KK^{*****}& 12.92 & 12.89\\ \hline
\end{array}
\]
\caption{The masses of the first few KK modes on the circle
in GeV. Along a generic direction these KK modes decouple, thus we have
chosen
an arbitrary point $a_1=2a_2=2.5 u_0$ for the comparison of the
numerical and WKB results. which are given in the first and second column.
\label{tab:kkcirc}}
\end{table}

\section{Conclusions}

In this paper we have presented a two-parameter
family of supergravity models of non-supersymmetric $3+1$ dimensional
$SU(N)$ Yang--Mills theory, based on
regular geometries with D4-brane charge. In these models, we have evaluated
the mass spectra of the scalar glueballs and some of the related
KK modes everywhere on the two dimensional parameter space
using both numerical and analytic (WKB) methods. We find that the glueball
mass ratios are very stable against the variation of the angular momentum
parameters.
The asymptotic values of these ratios  for large angular momenta
are in good agreement with the most recent lattice results everywhere
in the parameter space except along a special line $a_1=a_2\gg u_0$
(which is exactly the region where the WKB approximation breaks down, and
also the region where the inner and outer horizons approach each other and the
supergravity approximation approaches a discontinuous limit).
The KK modes on the compact D-brane coordinate
decouple for large angular momenta everywhere (except perhaps along
$a_1=a_2$, where our analysis of decoupling is inconclusive).
The KK modes on the $S^4$ however do not decouple
from the spectrum anywhere in the parameter space in the supergravity
approximation used in this paper.

The masses evaluated in this paper in the supergravity approximation
can in principle get large corrections when extrapolating
from the strong coupling (large $\lambda$) regime to the weak-coupling
regime of the Yang--Mills theory.
If the spectrum of the corresponding string models at small $\lambda $ do
indeed reproduce
the Yang--Mills spectrum, a natural question to ask
is why the glueball masses (or perhaps only the glueball mass ratios)
would get small corrections,
while the KK masses get large corrections.
Since in the limit $a_1,a_2\gg u_0$ the metric approaches the supersymmetric
space  (\ref{uuoo}),
it is possible that a subset of the masses may be protected by supersymmetry
(as explained in Sec.~2.3, the naive limit $u_0\to 0$
does not lead to a supersymmetric theory, since
it does not have fermions in the spectrum).
A problem of interest is thus to investigate the supersymmetric model
with $u_0=0$, and determine which scalars belong to short BPS multiplets,
and which ones are in long multiplets. Since the masses of the scalars
belonging to short multiplets should not be changed
in the $\lambda\ll 1$ regime, this could explain why
 the $0^{-+}$ glueball masses are so close to the lattice values, and it may
be used as a highly non-trivial quantitative test of the conjectured relation
of supergravity to non-supersymmetric $SU(N)$ Yang--Mills theory.

\setcounter{figure}{0}
\setcounter{table}{0}

\section*{Acknowledgements}

We thank Y. Oz
for useful discussions.
C.C. is a research fellow of the Miller
Institute for Basic Research in Science.
C.C. and J.T. are supported in part
the U.S. Department of Energy under Contract DE-AC03-76SF00098, and in part
by the National Science Foundation under grant PHY-95-14797.

\begingroup\raggedright\endgroup

\end{document}